\documentclass[a4paper]{jpconf}
\usepackage{graphicx}
\usepackage{amssymb,amsmath}

\newcommand{\C}{{\cal C}}

\newcommand{\J}{{\cal J}}

\begin{document}
\title{The sign problem and the Lefschetz thimble}

\author{Marco Cristoforetti, Luigi Scorzato\footnote{Speaker}}
\address{ECT$^\star$. Strada delle tabarelle, 286 -- I-38123, Trento, Italy.}

\author{Francesco Di Renzo} 

\address{Universit\`{a} di Parma and INFN gruppo collegato di Parma. Viale G.P. Usberti n.7/A -- I-43124, Parma,
  Italy.}

\begin{abstract}
Recently, we have introduced a novel approach to deal with the sign problem that prevents the Monte Carlo
simulations of a class of quantum field theories (QFTs).  The idea is to formulate the QFT on a Lefschetz thimble.
Here we review the formulation of our approach and describe the {\em Aurora} Monte Carlo algorithm that we are
currently testing on a scalar field theory with a sign problem.
\end{abstract}

\section{Introduction}

Recently, we have proposed a novel approach \cite{Cristoforetti:2012su} to deal with the sign problem that hinders
Monte Carlo simulations of many quantum field theories (QFTs).  The approach consists in formulating the QFT on a
Lefschetz thimble.  In this paper we concentrate on the application to a scalar field theory with a sign problem.
In particular, we review the formulation and the justification of the approach, and we also describe the {\em
  Aurora} Monte Carlo algorithm that we are currently testing.

\section{Formulation}

In this paper we consider a scalar field theory with $U(1)$ global symmetry:
\begin{equation}
\label{eq:Scont}
S= \int d^4 x 
[
|\partial \phi|^2
+
(m^2-\mu^2) |\phi|^2
+
\mu j_0
+ 
\lambda |\phi|^4
], 
\qquad\;
j_{\nu} := \phi^*\overleftrightarrow{\partial_{\nu}} \phi,
\end{equation}
where $\phi(x)$ is a complex scalar field. This model can be regularized on a lattice as:
\[
S[\phi]=\sum_x\left[
\left(2d+m^2\right)\phi_x^*\phi_x
+ \lambda(\phi^*_x\phi_x)^2
- \sum_{\nu=0}^{d-1}
  \left(
    \phi^*_x e^{-\mu\delta_{\nu,0}} \phi_{x+\hat{\nu}}
    + \phi^*_{x+\hat{\nu}} e^{\mu\delta_{\nu,0}} \phi_x
  \right)
\right],
\]
and has a sign problem when $\mu\neq 0$, which has been successfully treated in \cite{Endres:2006xu} and also in
\cite{Aarts:2008wh}.

Our approach consists in defining the observables as:
\begin{equation}
\label{eq:Z}
\langle {\cal O} \rangle_0 = \frac{1}{Z_0} \int_{\J_0} \; \prod_{a,x} d\phi_{a,x} \; e^{-S[\phi]} {\cal O}[\phi],
\qquad
Z_0 = \int_{\J_0} \; \prod_{a,x} d\phi_{a,x} \; e^{-S[\phi]},
\end{equation}
where the integration domain $\J_0$ is the Lefschetz thimble \cite{Pham1983,Witten:2010cx} attached to a global
minimum $\phi_{\rm glob}$ of the real part of the action $S_R=\Re{S}$, when restricted to the original domain
$\C=\mathbb{R}^{2V}$.  More precisely, $\J_0$ is the manifold of real dimension $n=2V$, defined as union of all the
curves of steepest descent (SD) for $S_R$, i.e., solutions of:
\begin{eqnarray}
\label{eq:SD}
\frac{d}{d\tau} \phi^{(R)}_{a,x}(\tau) &=& - \frac{\delta S_R[\phi(\tau)]}{\delta \phi^{(R)}_{a,x}},
\;\;\;\; \forall a,x,\\
\nonumber
\frac{d}{d\tau} \phi^{(I)}_{a,x}(\tau) &=& - \frac{\delta S_R[\phi(\tau)]}{\delta \phi^{(I)}_{a,x}},
\;\;\;\; \forall a,x,
\end{eqnarray}
that end in $\phi_{\rm glob}$ for $\tau\rightarrow\infty$.  Here we have assumed the usual complexification of the
complex scalar field (see, e.g., \cite{Aarts:2008wh}), where both the real ($\phi_1$) and imaginary ($\phi_2$) part
of the original fields become complex $\phi_a = \phi^{(R)}_a + i \phi^{(I)}_a$, $a=1,2$.

\subsection{Formulation in presence of SSB}

Note that the formulation described above is possible only when $\phi_{\rm glob}$ is a minimum of the action with
positive definite Hessian.  In presence of symmetries that act non-trivially in $\phi_{\rm glob}$ this is not the
case.

Gauge symmetries can be treated as described in \cite{AtiyahBott-YM, Witten:2010cx, Cristoforetti:2012su}, where,
essentially, the thimble is defined modulo gauge transformations.  But this is not suitable to study the
possibility of spontaneous symmetry breaking (SSB).  However, the proper way to study SSB (even in the ordinary
formulation of the functional integral) is to introduce a small term of explicit symmetry breaking and study the
limit when such term goes to zero.  In this way, $\phi_{\rm glob}$ becomes a true minimum and the thimble can be
defined.

\subsection{Justification of the approach}

The functional integral in Eq.~(\ref{eq:Z}) does not coincide, in general, with the standard formulation.  Morse
theory \cite{Witten:2010cx} only enables us to say that---under suitable conditions on $S[\phi]$ and ${\cal O}$---
the standard functional integral coincides with an integral over a particular combination of thimbles
$\sum_{\sigma} n_{\sigma} \J_{\sigma}$, for some integer $n_{\sigma}$, and where the sum runs over all the
stationary points of the complexified action\footnote{An argument in \cite{Witten:2010cx} (see Sec.~3.2), can be
  used to show that only the stationary points near to the global minima in $\C$ can bring a non negligible
  contribution.  See \cite{Cristoforetti:2012su}, Sec.~II.B.3.}.

However, we have shown in \cite{Cristoforetti:2012su} that the model defined by Eq.~(\ref{eq:Z}) has the same
degrees of freedoms, the same symmetries and symmetry representations, and also the same perturbative expansion as
the standard formulation.  In this sense, by universality, the formulation of Eq.~(\ref{eq:Z}) may be seen as an
alternative regularization of the model in Eq.~(\ref{eq:Scont}).

\section{The {\em Aurora} algorithm}

In this section we describe an updated form of the algorithm proposed in \cite{Cristoforetti:2012su}. 
What we want to compute are expectation values like:
\begin{equation}
\label{eq:ZR}
\frac{1}{Z_0}
e^{-i S_I} \int_{\J_0} \; \prod_x d\phi_x \; e^{-S_R[\phi]} {\cal O}[\phi],
\end{equation}
where the imaginary part of the action $S_I$ is constant in $\J_0$ and can be taken out of the integral.  Hence the
phase of $e^{-S}$ cancels from all observables.  Moreover, the real part $S_R$ is bounded from below by its value
at the stationary point\footnote{The integral is convergent, because $S[\phi]$ is a polynomial in $\phi$ and $\J_0$
  is union of the curves of SD.}.  The functional integral in Eq.~(\ref{eq:ZR}) possesses a bounded real action and
hence can be studied, e.g., with a Langevin algorithm\footnote{The phase that comes from the measure $d\phi$ will
  be treated with reweighting, as discussed in Sec.~\ref{sec:phase}.}.  The choice of the Langevin algorithm is
particularly convenient because the drift term of the Langevin equation for the real part of the action coincides
with Eq.~(\ref{eq:SD}), i.e., the SD.  Hence, such drift preserves the manifold $\J_0$ by construction.  The
Langevin noise, however, needs to be projected on the tangent space to $\J_0$, in order to explore the integration
domain correctly.

The projection on the tangent space of $\J_0$ at $\phi$ is challenging, because we seem to have no way to tell
which configurations in the neighborhood of $\phi\in \J_0$ will eventually also fall in $\phi_{\rm glob}$, by
following the SD flow.  However, the tangent space at the stationary point $\phi_{\rm glob}$ can be computed, by
computing the Hessian matrix\footnote{Which is easy to compute also analytically, when $\phi_{\rm glob}=0$, and can
  be computed numerically, once and for all, for general $\phi_{\rm glob}$.}.  In order to use this fact to
generate a noise vector tangent to $\J_0$ at $\phi$, we need a way to transport a vector $\eta$ along the flow of
SD, while keeping it tangent to $\J_0$.  One way to achieve this is by requesting that the vector $\eta$ is
parallel transported along the flow of $S_R$, i.e. requesting that its Lie derivative along $\partial S_R$ is zero:
\[
{\cal L}_{\partial S_R} (\eta)= [\partial S_R,\eta]=0
\]
which nicely translates into a first order ordinary differential equation, which is also linear in
$\eta$:
\begin{equation}
\label{eq:Lie}
\frac{d}{d\tau} \eta_{j}(\tau) = \sum_{k} \eta_{k}(\tau) \partial_{k}\partial_{j} S_R.
\end{equation}
Note that we use $j,k$ as multi-indices for $(R/I,a,x)$.

Actually, we can do better. Eq.~(\ref{eq:Lie}) ensures a parallel transport of $\eta$ along the flow $\partial
S_R$.  This is actually more than what we need, because we only need to keep $\eta$ tangent to $\J_0$.  On the
other hand, it would be nice to transport $\eta$ by means of an orthogonal flow, that preserves orthonormality,
isotropy, and also enables the use of stable numerical integrators (see Sec.~5.3 of \cite{iserles2008first}).  This
can be done quite easily by employing the Iwasawa decomposition of $\partial^2 S_R$ (see, e.g., Sec.~VI of
\cite{knapp2002lie}), which, at the group level, is a generalization of the Gram-Schmidt orthonormalization
procedure, and it can be implemented also at the algebra level.  More precisely, if we write:
\[
\left ( \partial^2 S_R \right)_{i,j} = A_{i,j} + D_{i,j} + N_{i,j},
\]
where $A$ is skew-symmetric, $D$ is diagonal and $N$ is upper triangular, then the flow
\begin{equation}
\label{eq:Iwa}
\frac{d}{d\tau} \eta(\tau) =  \eta(\tau) A(\tau)
\end{equation}
defines an orthogonal flow whose solutions $\eta$ span the same space spanned by the solutions of
Eq.~(\ref{eq:Lie}).  This means that the tangent space of $\J_0$ is still preserved.  But now, the flow defined by
Eq.~(\ref{eq:Iwa}) also preserves orthonormality, isotropy, and such properties can be even ensured numerically by
using suitable numerical integrators, such as the implicit midpoint rule \cite{iserles2008first}.

We can now summarize the algorithm as follows.  We use $t$ to represent the Langevin (Monte Carlo) time and $\tau$
to represent the parameter of the SD.  Let us assume that we have already a configuration $\phi_t$ in $\J_0$, and
we also have a path $\phi_t(\tau)$ that fulfills the equations of SD and connects $\phi_t$ to a configuration
$\phi_t^{(\varepsilon)}$ with norm less than $\varepsilon$.  The value of $\varepsilon$ must be sufficiently small
so that the quadratic approximation of the action is valid.  One step of Langevin proceeds as follows:
\begin{itemize}
\item apply the Langevin force, that consists in taking one step forward along the curve of SD which is already
  available.  The result is the configuration $\phi'_t$;
\item extract a Gaussian noise $\eta$;
\item project out the components of $\eta$ orthogonal to $\J_0$ at the origin.  More precisely, set
  $\eta_{\parallel}=(P-1)\eta$, where $P=\partial^2 S_R / \sqrt{(\partial^2 S_R)}$, and then re-scale the norm of
  $\eta_{\parallel}$ according to the $\chi$ distribution;
\item evolve $\eta_{\parallel}$ with Eq.~(\ref{eq:Iwa}) from $\phi_t^{(\varepsilon)}$ to $\phi'_t$.  The orthogonal
  properties of the flow can also be preserved numerically by employing, e.g., the implicit midpoint rule.  This
  leads to solve a 5-D linear system in $\eta_{\parallel}(\tau)$;
\item define a tentative new configuration as $\phi_{t+dt}^{\rm guess}= \phi'_t + \sqrt{2t}\, \eta(\tau)$;
\item use $\phi_{t+dt}^{\rm guess}$ as a starting point to find a solution of the SD equation.  This can also be
  integrated with the implicit midpoint rule with the constraint that $(P-1)\phi_{t+dt}^{(\varepsilon)}=0$.  This
  leads to a non-linear 5-D system that can be solved---if the initial guess is sufficiently good---with a number
  of Newton-Raphson iterations.
\end{itemize}

\subsection{The residual phase}
\label{sec:phase}

As noted in the previous section, the integral of Eq.~(\ref{eq:ZR}) also includes a residual phase that comes from
the determinant of the tangent space of $\J_0$ at $\phi$, and is not necessarily real and positive. We argue that
this should not represent a {\em sign problem}.  In fact, a sign problem means that the average phase $|\langle
\prod d\phi \rangle| \ll 1$.  However, that phase is constant in the region where the integrand is not very small.
In fact, such phase is completely neglected in the saddle point expansion, which, on the other hand, can be
identified with some form of perturbative expansion of the QFT.  Saying that the average phase could be negligible
is equivalent to saying that the ``perturbative contribution'' (in some sense that we cannot define precisely)
could be negligible.  We do expect that a non-perturbative contribution may be important, but that the perturbative
one could be negligible is contrary to the general expectations.

In this sense, we have good reasons to hope that residual phase can be treated with reweighting, but we have no
justification to neglect it.  If so, we have to compute it, which is quite expensive.  This can be done as follow.
If we call $T_{\phi}$ the complex ($2V\times 2V$) matrix that spans the tangent space of $\J_0$ at $\phi$, we have:
\[
\det(T_{\phi_{\tau}})
=
-\int_{\tau}^{\infty} 
ds \,
\mbox{Tr}
[\, T_{\phi_{s}}^{-1} \, \frac{d}{ds} T_{\phi_{s}}\, ]
\simeq
\sum_{i=0}^{N_{\tau}}
\mbox{Tr}
[\, T_{\phi_{i}}^{-1} \, \Delta T_{\phi_{i}}\, ]
=
\sum_{i=0}^{N_{\tau}}
\,
\sum_{k=1}^{N_R}
\,
{\xi^{(k)}}^T T_{\phi_{i}}^{-1} \, \Delta T_{\phi_{i}} \, \xi^{(k)}
\]
where the $\xi^{(k)}$, $k=1,\ldots,N_R$ are noisy estimators for the trace.  Assuming that the inversion of
$T_{\phi}$ would require $N^{CG}$ iterations, the computation of the determinant would cost as much as evolving
$N_R N^{CG}$ vectors, which is described in the previous section.

Some saving may also come from the fact that the computation of the phase is necessary only for the configurations
that are actually used for measurement.

\subsection{Preliminary tests}

\begin{figure}[h]
\begin{minipage}{18pc}
\includegraphics[width=18pc]{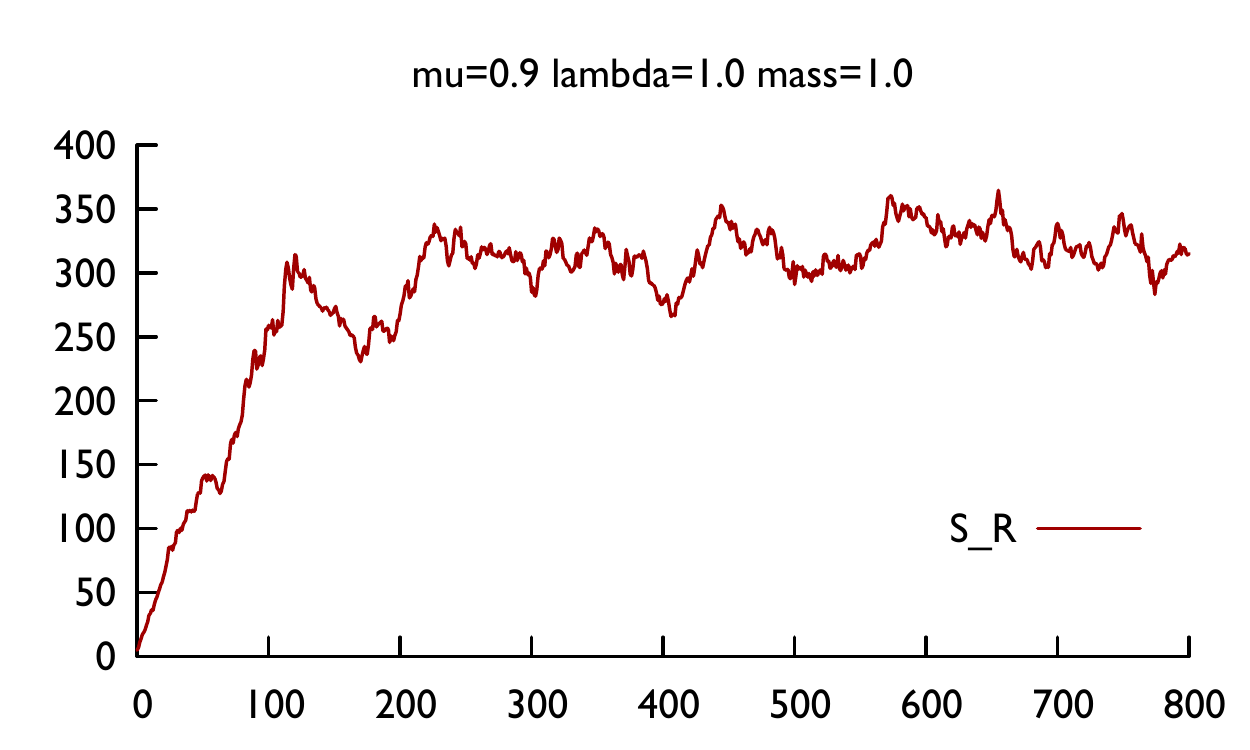}
\caption{\label{fig:1}\small The real part of the action on a $V=4^4$ lattice with Euler integrator. The system is
  stable long enough to show signs of a thermalization.}
\end{minipage}\hspace{2pc}%
\begin{minipage}{18pc}
\includegraphics[width=18pc]{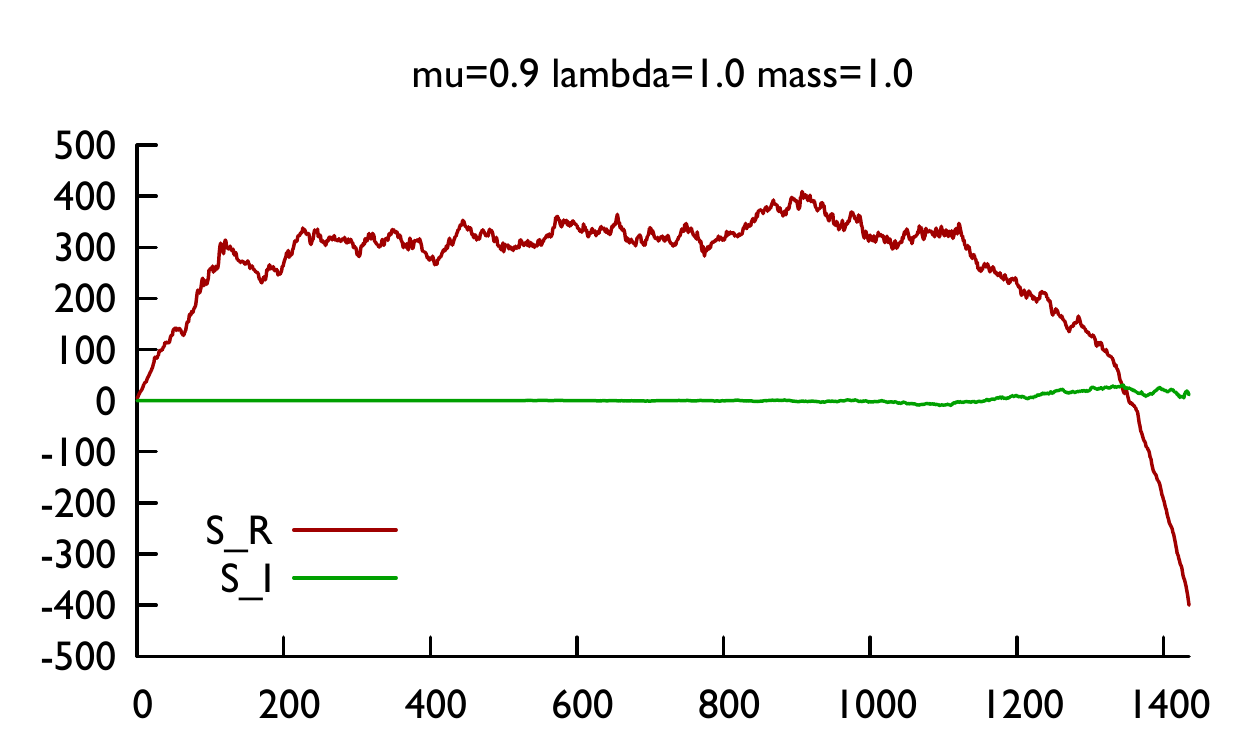}
\caption{\label{fig:2}\small Same as Fig.~\ref{fig:1}, but for a longer time.  The Euler integrator cannot
  constrain the system on the thimble, and the system drifts away at some point.  This is also confirmed by the
  deviation of the imaginary part of the action from a constant.}
\end{minipage} 
\end{figure}

As a first test, we implemented the method by using the naive Euler integration method.  Although bound to fail
because of the nature of the Euler method, this is an interesting test in order to get a feeling of how difficult
it is to keep the system on a thimble and how sensitive the system is to perturbations.  Some results are shown in
Fig.~\ref{fig:1} and \ref{fig:2}.  These are encouraging, because it seems that even the Euler method is able to
keep the system on the thimble long enough to see a thermalization.  This lets us hope that a better integrator
will have good chances to converge.  Observables appear around the expected order of magnitude, but their
thermalization is not clear.

\section{Conclusions}

We have illustrated an new approach to deal with the sign problem that afflicts a class of QFTs.  It consists in
regularizing the QFT on a Lefschetz thimble. Although it does not coincide with the usual regularization, it is a
legitimate one on the basis of universality. In fact, we could prove that QCD on the thimble has the same
symmetries, d.o.f. and PT as the usual formulation.  

We have also introduced a Monte Carlo algorithm to achieve an importance sampling of the configurations on the
thimble. Its numerical implementation will be certainly challenging and expensive, but all the steps of the
algorithm are, a priori, feasible and have acceptable scaling.  The residual phase should not give a sign problem
(unless we believe that the perturbative contribution can be negligible) and hence should be manageable with
reweighting, but this must be checked.  We are presently testing the method for a scalar QFT on tiny lattices with
encouraging results.

\ack 
This research is supported by the AuroraScience project (funded by the Provincia Autonoma di Trento and INFN),
and by the Research Executive Agency (REA) of the European Union under Grant Agreement No.  PITN-GA-2009-238353
(ITN STRONGnet).  L.S. and M.C. are members of LISC.  FDR is partially supported by INFN i.s. MI11 and by MIUR
contract PRIN2009 (20093BMNPR\_004).

\section*{References}
\bibliography{density}{}
\bibliographystyle{iopart-num}

\end{document}